\documentstyle[ltwol,epsfig]{article}

\def\Journal#1#2#3#4{{#1} {\bf #2}, #3 (#4)}

\def\NPB{{\em Nucl. Phys.} B}
\def\PLB{{\em Phys. Lett.}  B}
\def\PRL{\em Phys. Rev. Lett.}
\def\PRD{{\em Phys. Rev.} D}

\def\be{\begin{equation}}
\def\ee{\end{equation}}
\def\bea{\begin{eqnarray}}
\def\eea{\end{eqnarray}}

\begin{document}

\title{ \hfill OKHEP--98-06\\ANALYTIC PERTURBATIVE APPROACH TO 
QCD}

\author{K. A. MILTON}

\address{Department of Physics and Astronomy, The University of Oklahoma,
Norman, OK 73019-0225 USA \\E-mail: milton@mail.nhn.ou.edu}

\author{I. L. SOLOVTSOV, O. P. SOLOVTSOVA}

\address{Bogoliubov Laboratory of Theoretical Physics, Joint Institute
for Nuclear Research, Dubna, 141980 Russia \\E-mail:
solovtso@thsun1.jinr.ru, olsol@thsun1.jinr.ru}


\twocolumn[\maketitle\abstracts{A technique called
analytic perturbation theory, which respects
the required analytic properties, consistent with causality, is
applied to the definition of the running coupling in the timelike region,
to the description of inclusive $\tau$-decay, to deep-inelastic scattering sum
rules, and to the investigation of the renormalization scheme ambiguity. It is
shown that in the region of a few GeV the results are rather
different from those obtained in the ordinary perturbative description
and are practically renormalization scheme independent.}]

\section{ Analytic Running Coupling Constant}
The conventional renormalization-group resummation
\footnotetext{Presented at ICHEP'98, Vancouver, July 1998}
of perturbative series leads to unphysical
singularities in the running coupling constant. For example,
the usual QCD one-loop running coupling is~\footnote{We use the notation
$Q^2=-q^2$, where $Q^2>0$ for spacelike momentum transfer.}
\be
{\alpha^{\rm PT}(Q^2)\over4\pi}={1\over\beta_0}{1\over
\ln(Q^2/\Lambda^2)},
\ee
where $\beta_0=11-2n/3$, $n$ being the number of
quark flavors.  This evidently has a singularity (Landau pole) at
$Q^2=\Lambda^2$, which is unphysical and inconsistent with the causality
principle. Instead, we propose replacing perturbation theory (PT) by
analytic perturbation theory (APT)~\cite{shirkov,ms}
to enforce the correct analytic properties, for example, that the running
coupling be regular except for a branch cut for $-Q^2\ge0$, by adopting the
dispersion relation
\be
{\alpha^{\rm APT}(Q^2)\over4\pi}
={1\over\pi}\int_0^\infty d\sigma{\rho(\sigma)\over
\sigma+Q^2-i\epsilon} \, ,
\ee
where the imaginary part is given by the perturbative result, that is
\be
\rho(\sigma)={1\over 4\pi}\rm{Im}\,\alpha^{\rm PT}(-\sigma-i\epsilon)\, .
\ee
This leads to a consistent spacelike coupling, which in one-loop is:
\be
\alpha^{\rm APT}(Q^2)={4\pi\over\beta_0}\left[{1\over\ln(Q^2/\Lambda^2)}
+{\Lambda^2\over\Lambda^2-Q^2}\right] \, .
\ee
The second, nonperturbative term, cancels the ghost pole.

The above defines the running coupling in the spacelike region.
We can also define a timelike (or $s$-channel) coupling
$\alpha_s(s)$,\cite{ms}  which is related to the spacelike coupling
through the following reciprocal relations:
\begin{eqnarray}
\alpha_s(s)&=&-{1\over2\pi i}\int_{s-i\epsilon}^{s+i\epsilon} {dz\over z}
\alpha(-z) \, ,\\
\alpha(Q^2)&=&Q^2\int_0^\infty {ds\over(s+Q^2)^2} \alpha_s(s) \, ,
\end{eqnarray}
where the contour in the first integral does not cross the cut on the
positive real axis.  In terms of the spectral function, then,
\be
{\alpha_s(s)\over4\pi}
={1\over\pi}\int_s^\infty {d\sigma\over\sigma}\rho(\sigma)\, .
\end{equation}

\begin{figure}[t]
\centerline{
\epsfig{file=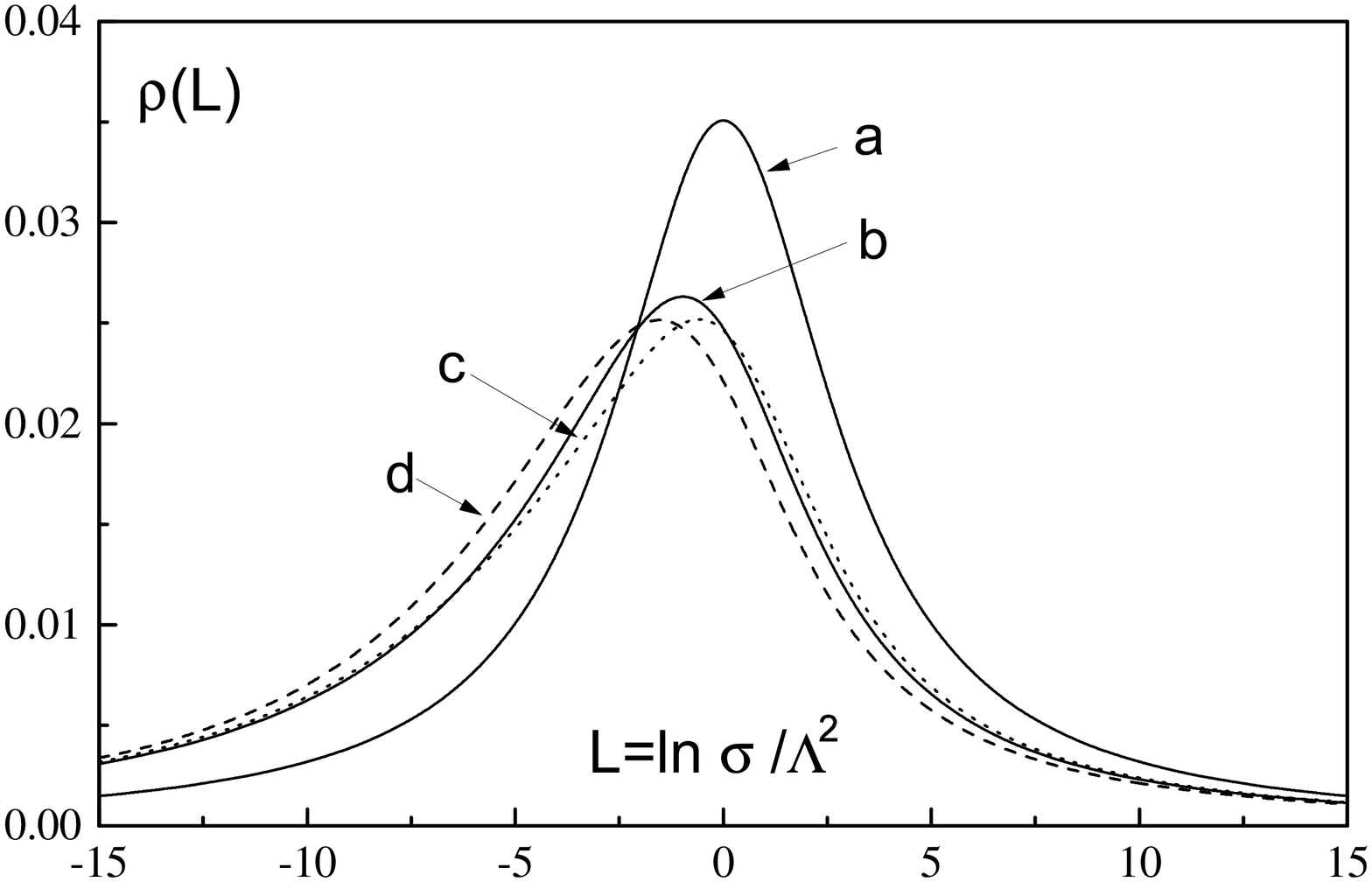,width=9.0cm}}
\caption{The spectral functions: (a) and (b) are the one- and two-loop
results; (c) is the two-loop approximation obtained as the first iteration of
the exact renormalization-group equation; \protect\cite{mss}
(d) is the three-loop result in the $\overline{\rm MS}$ scheme. }
\label{figsp}
\end{figure}

\begin{figure}
\centerline{\epsfig{file=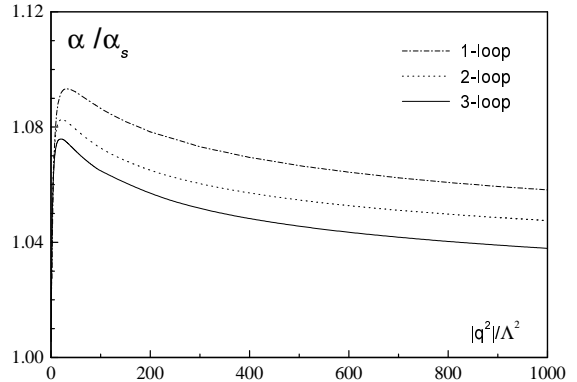,width=9cm}}
\caption{The ratio of spacelike and timelike values of the running
coupling constant in the APT approach, at one, two, and three loops.}
\label{fig1}
\end{figure}

The spectral functions in 1-, 2-, and 3-loops are
shown in Fig.~\ref{figsp}, for $n=3$.
The areas under all these curves
turn out to be the same, which implies a universal infrared fixed
point $\alpha^*$,
\be
\label{fixed_point}
{\alpha^*\over4\pi}={\alpha(0)\over4\pi}={\alpha_s(0)\over4\pi}
={1\over\pi}\int_0^\infty {d\sigma\over\sigma}\rho(\sigma)={1\over
\beta_0}\, ,
\ee
which is exact to all orders.~\cite{shirkov,mss}
It is also possible to prove that symmetrical behavior of the timelike
and spacelike couplings is inconsistent with the required analytic properties,
that is, in any renormalization scheme
\be
 \alpha_s(s)\ne \alpha(Q^2), \quad s=Q^2.
\ee
This difference, which is important when the value of the running coupling
is extracted from various experimental data,~\cite{msa}
is demonstrated in Fig.~\ref{fig1} for the $\overline{\rm MS}$ scheme.

\begin{figure}[thp]
\centerline{\epsfig{file=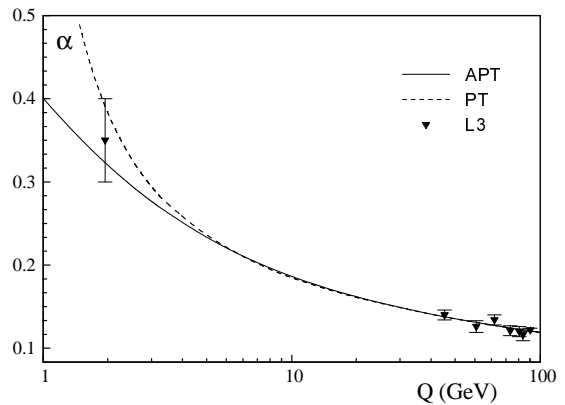,width=9cm}}
\caption{ QCD evolution of the running coupling constants
(defined in the spacelike region) compared to experimental
data.\protect\cite{L3}
Here a matching procedure across the various quark thresholds
has been applied in the timelike region.\protect\cite{msa}
}
       	\label{fig3}
\end{figure}

Because of the infrared fixed point, the running coupling implied by APT
rises less rapidly for small $Q^2$ than does the usual perturbative running
coupling.  This is illustrated~\cite{msa} in Fig.~\ref{fig3}.

Another interesting fact is that within APT  Schwinger's
conjecture~\cite{js} about the connection between the $\beta$
and spectral functions is valid.\cite{ms}
Indeed, the $\beta$ function for the timelike coupling is
\be
\beta_s={s\over4\pi}{d \alpha_s\over ds}=-{1\over\pi}\rho(s).
\ee

\begin{figure}
\centerline{
\epsfig{file=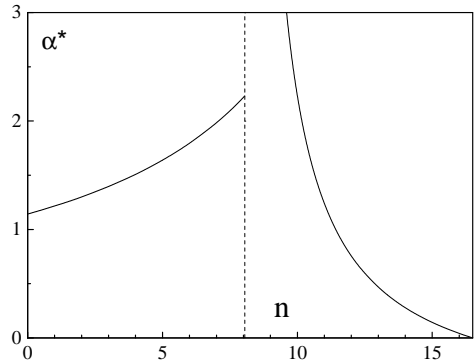,width=7.5cm}}
\caption{ The analytic infrared fixed point  $\alpha^*$
vs.~$n$.}
\label{fig:freeze}
\end{figure}

The above discussion assumes that the number of active flavors $n$ is
realistically small.  However, if
$n$ is large enough, even the perturbative coupling can exhibit an
infrared fixed point, at least in the two-loop level.  This same fixed
point defined by Eq.~(\ref{fixed_point}) occurs in the analytic approach
and we have
\begin{equation}
\frac{\alpha^{*}}{4\pi}=
\cases{
1 / \beta_0 \, , & $n \leq 8 \, , $  \cr
- \beta_0 / \beta_1 \, , &  $9 \leq n \leq 16 \, .$ \cr }
\end{equation}
The transition between the nonperturbative and perturbative fixed
point occurs for $n=8.05$, as is shown in Fig.~\ref{fig:freeze}.
This is qualitatively consistent with the phase transition seen, for
example, in nonperturbative approaches and lattice simulations.\cite{lat}

\vspace*{-1.8pt}   

\section{Inclusive $\tau$ Decay within APT}\label{sec:tau}
The inclusive semileptonic $\tau$ decay ratio for massless quarks
is given~\cite{Braaten} in terms of the electroweak
factor $S_{\rm EW}$, the CKM matrix
elements $V_{ud}$, $V_{us}$, and the QCD correction $\Delta_{\tau}$,
\be
R_\tau=3 S_{\rm EW}(|V_{ud}|^2+|V_{us}|^2)(1+\Delta_{\tau})\, .
\ee
The correction $\Delta_{\tau}$ may be written in terms of the functions
$r(s)$ or $d(q^2)$, which are the QCD corrections to the imaginary
part of the hadronic correlator $\Pi$: ${\rm Im}\,\Pi\sim1+r$ and to
the Adler $D$-function: $D=-q^2d\Pi/dq^2\sim 1+d$.
The analytic properties of the Adler function allow us to write down
the relations
\bea
d(q^2)&=&-q^2\int_0^\infty {ds\over (s-q^2)^2} r(s),\label{dapt}\\
r(s)&=&-{1\over2\pi i}\int_{s-i\epsilon}^{s+i\epsilon} {dz\over z}d(z).
\eea
Because of the proper analytic properties (which are violated in the
usual perturbative approach) the QCD contribution to the $\tau$ ratio
is given by the two equivalent forms
\bea
\Delta_{\tau}&=&2\int_0^{M_\tau^2} {ds\over M_\tau^2}
\left(1-{s\over M_\tau^2}
\right)^2\left(1+2{s\over M_\tau^2}\right) r(s)\label{cont}\\
&=&{1\over2\pi i}\oint_{|z|=M_\tau^2}{dz\over z}\left(1-{z\over M_\tau^2}
\right)^3\left(1+{z\over M_\tau^2}\right) d(z).\nonumber
\eea
If one likes, one can think of $r$ and $d$ as effective running couplings
in the timelike and spacelike regions, so as with the running couplings
they may be expressed in terms of an effective spectral density,
\bea
d^{\rm APT}(q^2)&=&{1\over\pi}\int_0^\infty {d\sigma\over\sigma-q^2}\rho^{\rm
eff}(\sigma),\label{drd}\\
r^{\rm APT}(s)&=&{1\over\pi}\int_{s}^\infty {d\sigma\over\sigma}\rho^{\rm eff}
(\sigma),
\eea
possessing the same universal infrared limit as the running coupling.

In the ordinary perturbative approach, the function $d(q^2)$ may be expanded
in terms of the running coupling and in the third order is
\be
\label{PT_exp}
d^{\rm PT}(q^2)=a^{\rm PT}(-q^2)+d_1 \left[a^{\rm PT}(-q^2)\right]^2
+d_2\left[a^{\rm PT}(-q^2)\right]^3,
\ee
where we have introduced $a=\alpha_S/\pi$, and, numerically, for three
active flavors, the coefficients are $d_1^{\overline{\rm MS}}=1.6398$,
$d_2^{\overline{\rm MS}}=6.3710$.
Such is not the case in APT; rather, $d^{\rm APT}$ is constructed from
Eq.~(\ref{drd}) with a spectral density obtained as the imaginary part of
$d^{\rm PT}$ on the physical cut:
\be
\rho^{\rm eff}(\sigma)=\rho_0(\sigma)+d_1\rho_1(\sigma)+d_2\rho_2(\sigma),
\ee
where $\rho_n(\sigma)={\rm Im}\,a^{n+1}(-\sigma-i\epsilon)$.
We use the world average \cite{PDG96} value\footnote{This is consistent
with the 1998 PDG value,\protect\cite{PDG98} which we extract as
$R_\tau=3.650\pm0.016.$} $R_\tau=3.633\pm0.031$.
Our results are shown in Table~\ref{tab1}, where for the sake of
illustration we also show the PT results obtained by using the contour
integral representation given in the second line of Eq.~(\ref{cont}).
\begin{table}
\begin{center}
\caption{The APT and PT parameters in the $\overline{\rm MS}$ scheme
extracted from $\tau$-decay.
The first two rows refer to NNLO calculations, the last two rows to
NLO calculations. The errors in the last digits are shown in
parentheses. }\label{tab1}
\vspace{0.2cm}
\begin{tabular}{|cccc|}
\hline
\raisebox{0pt}[12pt][6pt]{Method} &
\raisebox{0pt}[12pt][6pt]{$\Lambda_{\overline{\rm MS}}$ (MeV)} &
\raisebox{0pt}[12pt][6pt]{$\alpha(M_\tau^2)$} &
\raisebox{0pt}[12pt][6pt]{$d(M_\tau^2)$} \\
\hline
\raisebox{0pt}[12pt][6pt]{APT} &
\raisebox{0pt}[12pt][6pt]{871(155)} &
\raisebox{0pt}[12pt][6pt]{0.3962(298)} &
\raisebox{0pt}[12pt][6pt]{0.1446(88)} \\
\raisebox{0pt}[12pt][6pt]{PT} &
\raisebox{0pt}[12pt][6pt]{385(27)} &
\raisebox{0pt}[12pt][6pt]{0.3371(141)} &
\raisebox{0pt}[12pt][6pt]{0.1339(69)} \\
\raisebox{0pt}[12pt][6pt]{APT} &
\raisebox{0pt}[12pt][6pt]{918(151)} &
\raisebox{0pt}[12pt][6pt]{0.3983(236)} &
\raisebox{0pt}[12pt][6pt]{0.1431(84)} \\
\raisebox{0pt}[12pt][6pt]{PT} &
\raisebox{0pt}[12pt][6pt]{458(31)} &
\raisebox{0pt}[12pt][6pt]{0.3544(157)} &
\raisebox{0pt}[12pt][6pt]{0.1400(67)} \\
\hline
\end{tabular}
\end{center}
\end{table}
\vspace*{3pt}

\begin{figure}[thp]
~\vspace*{-0.5cm}
\centerline{
\epsfig{file=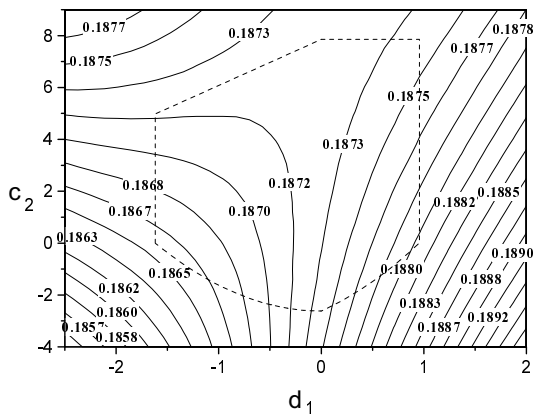,width=9.0cm}}
~\vspace*{0.2cm}
\caption{ Contour plot of the APT correction $\Delta_\tau$ at
	 the three-loop order as a function of RS parameters $d_1$\ and
	 $c_2$. The dashed line indicates the boundary of the domain
	 defined by Eq.~(\protect\ref{domain}).}
    \label{figrs}
\end{figure}

Most remarkably, APT exhibits very little renormalization scheme (RS)
dependence. The $\tau$ decay coefficients $d_1$ and $d_2$ are RS dependent,
as are all but the first two beta function coefficients, defined by the
renormalization group equation
\be \mu^2 {da\over d\mu^2}=-{b\over 2} a^2(1+c_1a+c_2 a^2) \,.
\ee
Because of the existence of RS invariants, one can investigate
the sensitivity of the predicted value of the QCD correction to the
choice of RS by varying $d_1$ and $c_2$
in a region where the degree of cancellation in the second RS invariant
$ \omega_2=c_2+d_2-c_1d_1-d_1^2 $
does not exceed a specified limit, taking, for example,
\be \label{domain}
{|c_2|+|d_2|+c_1|d_1|+d_1^2\over|\omega_2|}\le2 \, .
\ee
That sensitivity is shown \cite{msy} in Fig.~\ref{figrs}, based on
$\Delta_{\tau}^{\overline{\rm MS}}=0.1881$.
Observe that the relative difference between the prediction
of the lower corners of the domain defined by Eq.~(\ref{domain}) is 0.8\%,
while in PT that difference
is 5\%.  Note that the $\overline{\rm MS}$ scheme lies outside this
domain, as does the so-called V scheme, the prediction of
which differs by only 0.2\%
from the $\overline{\rm MS}$ value in APT, but by 66\% in PT!  To all
intents and purposes, APT exhibits practically no RS dependence.
(Because of the perturbative stability,
the known three-loop level is quite adequate for these RS recalculations.)

\section{Deep-Inelastic Scattering Sum Rules}
At present, the polarized Bjorken and  Gross--Llewellyn Smith
deep inelastic scattering sum rules allow the possibility,\cite{Albrow}  
as with $\tau$ decay,  of extracting the value of $\alpha_S$ from experimental 
data  at low $Q$, here down to $1\,{\mbox{\rm GeV}}$.
\subsection{Bjorken Sum Rule}
The Bjorken sum rule refers to the value of the integral of the difference
between the polarized structure functions of the neutron and proton,
\bea
\Gamma_1^{\rm p-n}
&=&\int_0^\infty dx\,\left[g_1^p(x,Q^2)-g_1^n(x,Q^2)\right]\\
&=&{1\over6}\left|g_A\over g_V\right|\left[1-\Delta_{\rm Bj}(Q^2)\right] \, ,
\eea
where the prefactor in the second line is the parton-level description.
In the conventional approach, with massless quarks, the QCD correction
is given by a power series similar to Eq.~(\ref{PT_exp}),
with coefficients for three active flavors being $d_1^{\overline{\rm MS}}
=3.5833$, $d_2^{\overline{\rm MS}}=20.2153$. However, this description
violates required analytic properties of the structure function moments,
which, as has been argued,\cite{Wetzel} follow from the existence
of the De\-ser-Gil\-bert-Sudar\-shan integral representation.
\begin{figure}[thp]
\centerline{\epsfig{file=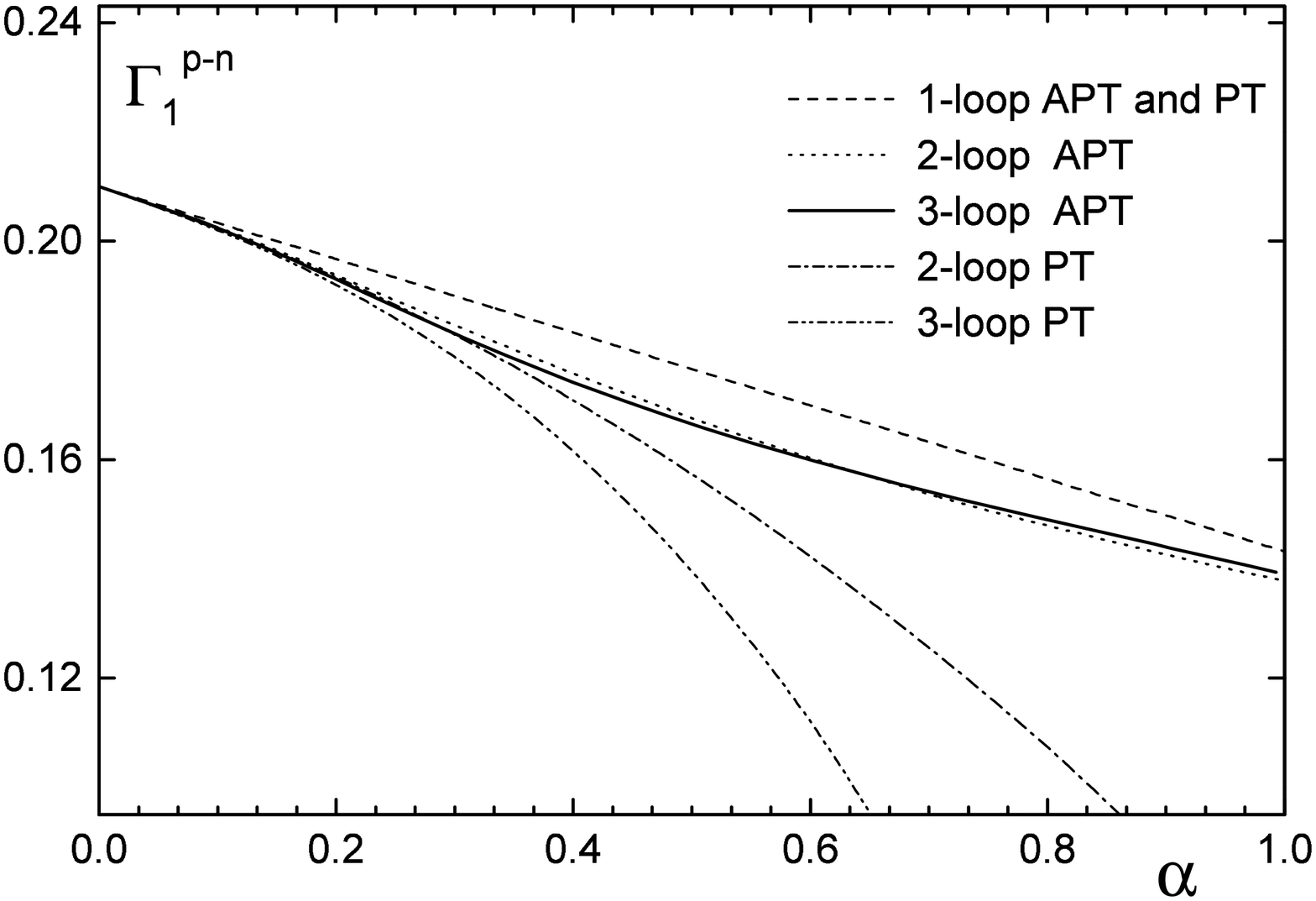,width=9.5cm}}
\caption{{ $\Gamma^{\rm p-n}_1\,$  with $1$-, $2$-, and $3$-loop
QCD corrections vs.\ the coupling constant. }}
\label{bj_fig1}
\end{figure}

Thus we adopt the analytic approach, which says instead
\be
\Delta_{\rm Bj}^{\rm APT}(Q^2)=\delta^{(1)}(Q^2)+d_1\delta^{(2)}(Q^2)
+d_2\delta^{(3)}(Q^2) \, ,
\ee
where
\be \delta^{(k)}(Q^2)=
{1\over\pi^{k+1}}\int_0^\infty {d\sigma\over\sigma+Q^2}
{\rm Im}\left[\alpha_{\rm PT}^k(-\sigma-i\epsilon)\right],
\ee
that is, $\Delta^{\rm APT}$ is not a power series in $\alpha^{\rm APT}$.

\begin{figure}[thp]
\centerline{
\hspace*{0.5cm}
\epsfig{file=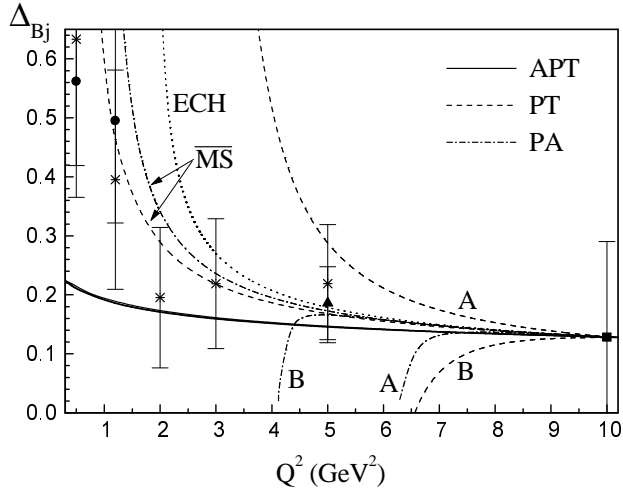,width=10.5cm}}
\caption{ Renormalization scheme dependence of predictions for
$\Delta_{\rm Bj}$ vs. $Q^2$ for the APT and PT expansions.
The solid curves, which are very close to
each other, correspond to the APT result
in the $\overline {\rm MS}$, A, B, and ECH schemes. The PT evolution
in the $\overline {\rm MS}$, A, and B schemes are shown by dashed curves,
the ECH scheme is indicated by a dotted curve, and
the PA results in the $\overline {\rm MS}$, A, and B schemes are denoted by
dash-dotted curves. (The definition of the various schemes is given in our
paper.\protect\cite{bj})
The SMC data~\protect\cite{smc} is indicated by a square, the
triangle is the E154 data,\protect\cite{e154}
circles are E143 data,\protect\cite{e143} and the stars are
recent E143 data.\protect\cite{newe143}
}
\label{bj_fig3}
\end{figure}

Besides possessing the correct analyticity, the APT approach has two
key properties in its favor.  First, successive perturbative corrections
are small, so that the two- and three-loop QCD results  are nearly the
same.  This is not the case in the conventional PT approach, as is
shown \cite{bj} in Fig.~\ref{bj_fig1}. Second, the renormalization scheme
dependence is again very small, so that various schemes which have the same
degree of cancellation as the $\overline{\rm MS}$ scheme give the same
predictions all the way down to $Q^2 \sim 1$ GeV$^2$. In contrast, it is
impossible to make any reliable prediction for conventional PT,
even if improved by the Pad\'e approximant (PA)
method,\cite{PA} for $Q^2$ below several GeV$^2$.
This is illustrated \cite{bj} in Fig.~\ref{bj_fig3}.              
One can see that instead of  RS unstable and rapidly changing PT functions,
the APT predictions are slowly varying functions, which
are practically RS independent.
\subsection{Gross-Llewellyn Smith Sum Rule}
A precisely similar analysis can be performed on the GLS sum rule,
which refers to the integral
\bea
S_{\rm GLS}&=&{1\over2}\int_0^1dx\left[F_3^{\overline\nu p}(x,Q^2)
+F_3^{\nu p}(x,Q^2)\right]\\
&=&3\left[1-\Delta_{\rm GLS}(Q^2)\right].
\eea
Again, the APT approach leads to perturbative stability, and to practically
no renormalization scheme dependence down to very low $Q^2$.\cite{gls}

\section{Conclusions}
Our conclusions are four-fold.
\begin{itemize}
\item APT maintains correct analytic properties (causality), and allows
for a consistent extrapolation between timelike ($\tau$ decay) and
spacelike (DIS sum rules) data.
\item Three loop corrections are much smaller than in PT; thus, there
is perturbative stability.

\item Renormalization scheme dependence is drastically reduced.
The three-loop APT level is practically RS independent.

\item The values of $\Lambda$ are larger in the APT approach than in
the PT approach.  Yet these values are consistent between timelike and
spacelike processes, and consistent with the data.
\end{itemize}

The work reported here is the beginning of a systematic attempt to
improve upon the results of perturbation theory in QCD. In the future
we will treat in detail the significance of power corrections
which come from the operator product expansion, and examine the effect
of finite mass corrections, which necessitate a more elaborate analytic
structure.

\section*{Acknowledgements}
This work was supported in part by grants from the US DOE, number
DE-FG-03-98ER41066, from the US NSF, grant number PHY-9600421, and from
the RFBR, grant 96-02-16126.
Useful conversations with D.~V.~Shirkov and L.~Gamberg
are gratefully acknowledged.
We dedicate this paper to the memory of our late colleague Mark Samuel.

\end{document}